\documentclass[preprint,amsmath,amssymb,superscriptaddress]{revtex4-1}

\usepackage{latexsym}        
\usepackage{graphicx}        
\usepackage{rotating}        
\usepackage{epstopdf}

\usepackage{dcolumn}
\usepackage{bm}

\begin{document}

\preprint{}

\title{Pre-pairing and the ``Filling" Gap in the Cuprates From the Tomographic Density of States}

\author{T. J. Reber}
\affiliation{Dept. of Physics, University of Colorado, Boulder, 80309-0390, USA
}%
\author{N. C. Plumb}
\altaffiliation{Now at Swiss Light Source, Paul Scherrer Institut, CH-5232 Villigen PSI, Switzerland
}%
\affiliation{Dept. of Physics, University of Colorado, Boulder, 80309-0390, USA
}%

\author{Y. Cao}
\author{Z. Sun}
\author{Q. Wang}
\author{K. P. McElroy}
\affiliation{Dept. of Physics, University of Colorado, Boulder, 80309-0390, USA
}%

\author{H. Iwasawa}
\author{M. Arita}
\affiliation{
Hiroshima Synchrotron Radiation Center, Hiroshima University, Higashi-Hiroshima, Hiroshima 739-0046, Japan
}%

\author{J. S. Wen}
\author{Z. J. Xu}
\author{G. Gu}
\affiliation{
Condensed Matter Physics and Materials Science Department, Brookhaven National Labs, Upton, New York,11973 USA
}%

\author{Y. Yoshida}
\author{H. Eisaki}
\author{Y. Aiura}
\affiliation{
AIST Tsukuba Central 2, 1-1-1 Umezono, Tsukuba, Ibaraki 3058568, Japan
}%

\author{D. S. Dessau}
\affiliation{Dept. of Physics, University of Colorado, Boulder, 80309-0390, USA
}%

\date{\today}

\begin{abstract}
We use the tomographic density of states (TDoS), which is a measure of the density of states for a single slice through the band structure of a solid, to study the temperature evolution of the superconducting gap in the cuprates.  The TDoS provides unprecedented accuracy in determining both the superconducting pair-forming strength, $\Delta$, and the pair-breaking rate, $\Gamma$.  In both optimally- and under-doped Bi$_2$Sr$_2$CaCu$_2$O$_{8+\delta}$, we find the near-nodal $\Delta$ smoothly evolves through the superconducting transition temperature - clear evidence for the existence of pre-formed pairs.  Additionally, we find the long observed `filling' of the superconducting gap in the cuprates is due to the strongly temperature dependent $\Gamma$.
\end{abstract}

\maketitle

The nature of the superconductive gap in the cuprates, the temperature at which it disappears, and how it disappears, are critical questions that have resisted explanation due to ambiguous spectroscopic measurements\cite{BasovRMP}\cite{DamascelliRMP}\cite{FischerRMP}.  At the antinode, where the gap is largest and most studies are focused, the ambiguity is unavoidable, because additional ordering potentially contaminates the region \cite{HanaguriCheckerboard}\cite{WiseCheckerboard} such that antinodal gaps may not be from the superconducting pairs.  The near-nodal region does not have these complications of additional ordering.  Unfortunately these near-nodal states have been barely explored spectroscopically.  In principle, the momentum resolution of angle resolved photoemission (ARPES) allows the near-nodal states to be studied independently from the antinodal gaps.  However, the limited experimental resolution of most ARPES experiments limits studies of the smaller near-nodal gaps, which we overcome with the meV-scale resolution of low-energy ($h\nu < 10 eV$) ARPES\cite{KoralekLaserARPES}.

Additionally, we show here that strongly temperature-dependent pair-breaking scattering processes causes the traditional ARPES-based methods of determining gap sizes to fail. This failure manifests when the pairing gap and the scattering energy scales are of comparable magnitude. We counter this problem with a new technique, the tomographic density of states (TDoS)\cite{ReberArcs}. Compared to regular ARPES, the TDoS method returns quantitatively more accurate values of the gap value $\Delta$ and dramatically smaller values (roughly an order of magnitude) for the electronic scattering rates $\Gamma$. For the first time, these values of the scattering rates qualitatively match what is observed from other probes such as optics and tunneling\cite{HwangOpticalScattering,AlldredgeGamma,PasupathyInhomogeneity}.  A comparison of our measured $\Delta$ and $\Gamma_{TDoS}$, highlight the strong interplay between these competing processes.   We find the near-nodal gap magnitude remains relatively constant through $T_C$, ($85-90K$ depending upon doping), and instead closes at a higher temperature $T_{Close}\approx$125K.  This new scale is well below the typical ``pseudogap" $T^*$ scale \cite{Huefner2Gap} but roughly consistent with the temperature scale at which Nernst \cite{WangNernst} and diamagnetism experiments \cite{LuLiDiamagnetism} show the onset of superconducting fluctuations. Our finding conclusively indicates that pre-formed Cooper pairs continue to exist in the normal state, whether or not an additional order parameter also contributes to pseudogap physics.  We also find that the energy gain for superconductive gapping primarily disappears due to the increasing $\Gamma$ filling in the gap, rather than the BCS case in which a shrinking $\Delta$ closes the gap.

In solids, the density of states represents the relative number of momentum states at a given energy over the whole Brillouin zone.  When the electron distribution is strongly momentum dependent, as in a d-wave superconductor, the density of states can be difficult to analyze.   The angular resolution of angle resolved photo-emission spectroscopy (ARPES) can select an individual slice of the Brillouin zone (Fig. 1a). By integrating along such a slice in momentum after removing the incoherent background, we extract a density of states for that slice, effectively a tomographic density of states (tomography is the imaging of a volume by individual slices). If the slice is chosen such that the momentum dependent variables are effectively single-valued, analysis is greatly simplified. Here we do that for the d-wave superconducting gap by taking slices roughly perpendicular to the Fermi surface.

To isolate the effect of the gap and remove the Fermi edge, we normalize the gapped off-nodal spectral weight to the gapless nodal spectral weight.  In this way we access some of the thermally populated states above $E_F$, which give the true (non-symmetrized) spectral function above $E_F$.  This curve is a TDoS, which is equivalent to a typical Giaever tunneling spectrum\cite{Giaever1960,Giaever1962}, except it is localized to single spot on the Fermi surface.   The TDoS is similar to other techniques that involve integration of ARPES spectra\cite{EvtushinskyIEDC,VekhterVarma}, but the isolation of the band and the normalization enable qualitative observations as well as simplify quantitative analysis.\footnote{See Supplemental Material at [URL will be inserted by publisher] for further technical details about the creation of the TDoS.}
\begin{figure}[htbp]
\includegraphics[width=90mm]{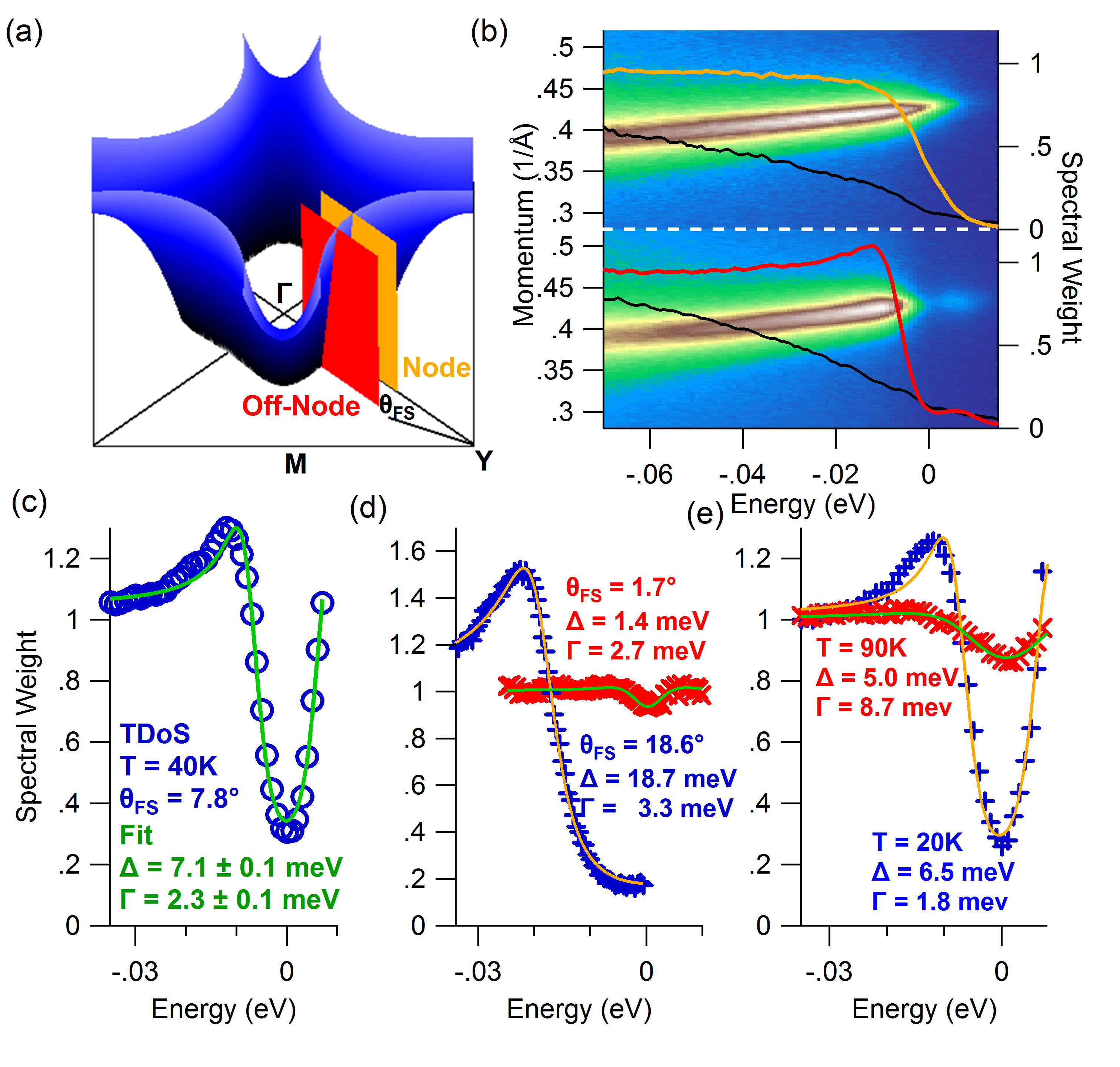}
\caption{\label{fig:Fig1}{\bf Creating and Fitting the Tomographic Density of States (TDoS)} {\bf (a)}) Tight-binding model of the Bi2212 band structure slices of which ARPES samples.  A nodal slice (orange) and an off-nodal slice (red) are shown.  The location of each slice is characterized by the Fermi surface angle, $\theta_{FS}$, measured from the node about the Y-point {\bf (b)} Nodal and off-nodal ARPES spectra  with corresponding coherent spectral weights (color) and incoherent backgrounds (black) from an optimally doped ($T_C$=91K) Bi2212 sample at T=40K. {\bf (c)}  Tomographic density of states created by normalizing the off-nodal spectral weight by the nodal spectral weight as well as the fit to the Dynes formula {\bf (d)} Two TDoS (T=50K) and the respective fits for T=50K and angles ranging from less than two degrees from the node to 40\% of the way to the anti-node. {\bf (e)} Two TDoS from $\theta_{FS}=7.8^\circ$and the respective fits for temperatures ranging from deep in the superconducting state until just below $T_C$.}
\label{fig1}
\end{figure}
To analyze the TDoS, we use the formula first proposed by Dynes to explain tunneling spectra from strongly coupled s-wave superconductors\cite{DynesFormula}:
\begin{equation} \label{eq:DynesFormula}
\rho_{Dynes}=Re\frac{\omega-i\Gamma}{\sqrt{(\omega-i\Gamma)^2-\Delta^2}}
\end{equation}
where $\omega$ is the energy relative to $E_F$,  $\Gamma$ is the  pair-breaking rate\cite{Mikhailovsky1991511} and $\Delta$ is the superconducting gap. Since each TDoS is specific to a location on the Fermi surface the gap strength is single valued, allowing us to use Dynes's original form even in d-wave superconductors.    The Dynes formula fits TDoS curves over a wide range of angles (Fig. 1d) and temperatures (Fig. 1e), for the material studied in this letter, Bi$_2$Sr$_2$CaCu$_2$O$_{8+\delta}$ (Bi2212).  The accuracy of the fits suggests the Dynes formula is a good model for the underlying physics probed by the TDoS.  For the slightly off-nodal TDoS of Fig. 1c the fit returns a local gap magnitude, $\Delta$ of 7.1 $\pm$ 0.1 meV and the pair-breaking rate, $\Gamma_{TDoS}$, of 2.3 $\pm$ 0.1 meV (Fig. 1c). This pair-breaking rate is roughly an order of magnitude smaller than the full electron scattering rate that ARPES measures through the EDC or MDC widths ($\approx15-20$ meV at $E_F$ for the samples measured in this paper).\footnote{See Supplemental Material at [URL will be inserted by publisher] for further technical details about the fitting procedure.}

We attribute the difference between $\Gamma_{TDoS}$ and $\Gamma_{MDC}$ to forward scattering events - the (impurity-based) forward scattering processes that shift an initial electron $k$ to a final $k'$.  In contrast, self-energy processes that appear in the spectral function are those that have a virtual intermediate state $k'$, but both initial and final states $k$.  These self-energy processes renormalize the band dispersion and are strongly energy dependent, while the impurity scattering events are energy independent and do not alter the $E$ vs $k$ dispersion.  Consequently, they should be added in to the spectral function separately.  More specifically, large angle impurity scattering (e. g. Cu vacancies) is likely to generate a weak background of states at all $k$ values, which can be added to the spectral function as a constant offset.  Small angle scattering, (e. g. out-of-plane disorder) will have the effect of a small amount of momentum broadening of the spectral function, which can be approximated by convolving the spectral function with a Lorentzian in the momentum direction. Alternatively, a self-energy broadening has a fundamental scale that is the lifetime of the excited state, converting to a broadening of the underlying band in energy.  For a system with a linear dispersion, it is very difficult to distinguish between self-energy broadening and forward scattering broadening.  However when the band curves (e. g. the bent back band from the gap formation) momentum broadening will not shift weight above the top of the band, while the self-energy broadening will, so the two effects become separable.  The integration at the heart of the TDoS distinguishes the two, because forward scattering broadens a band in momentum, but neither creates nor destroys spectral weight at a particular energy as measured by the TDoS. Consequently, the integrated TDoS is only sensitive to the self-energy terms.
\begin{figure}[htbp]
\includegraphics[width=90mm]{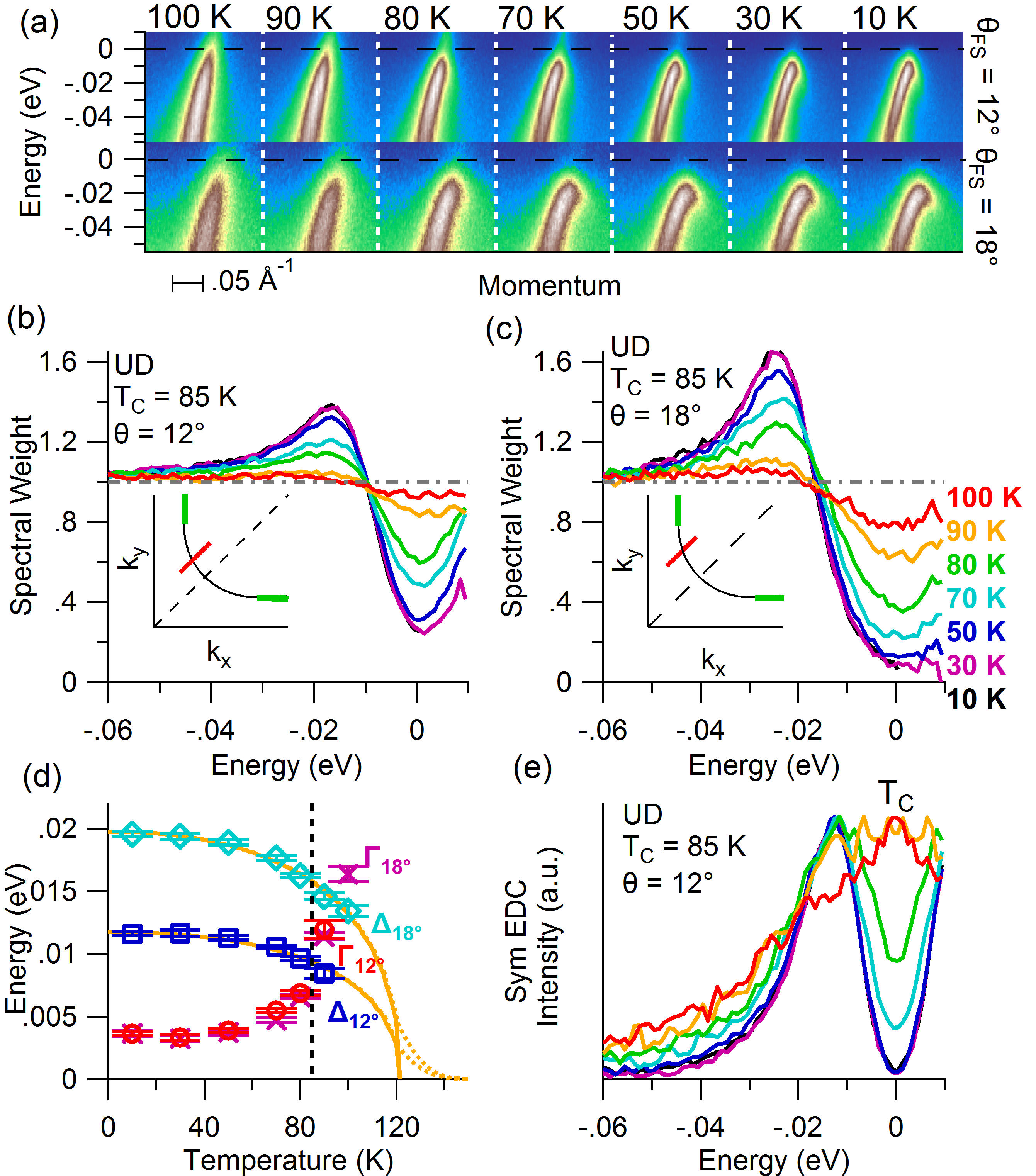}
\caption{\label{fig:Fig2}{\bf Temperature Dependence of the TDoS} {\bf (a)} Temperature evolution of ARPES spectra for two angles in the near nodal region for lightly underdoped ($T_C$=85K) Bi2212.  {\bf (b)} Temperature evolution for TDoS for 12$^\circ$ from the node (marked in red on inset Brillouin zone the anti-nodal pseudogap region\cite{TanakaTwoGap} is marked in green).  The inset shows a zoom of the 90K TDoS (orange) and its fit to the Dynes formula(black)  {\bf (c)} Temperature evolution for TDoS for 18$^\circ$ from the node The inset shows a zoom of the 100K TDoS (red) and its fit to the Dynes formula(black) {\bf (d)} Fit results of TDoS to Dynes formula showing a strongly temperature dependent $\Gamma$ and slowly evolving $\Delta$, which continues to exist above $T_C$. The error bars represent twice the statistical uncertainties, as returned from our least-squares fitting procedures, to incorporate systematic errors as well ($\pm 2\sigma$). {\bf (e)} Temperature evolution of symmetrized EDCs for $\theta_{FS}=12^\circ$ showing the formation of a resolvable peak away from $E_F$ only exists below $T_C$ (the same color scale for temperature is used in panels b,c, and e). The EDCs have been rescaled to emphasize the evolution of the lineshape.}
\label{fig2}
\end{figure}

In Fig. 2, we show the temperature evolution of the TDoS through the superconducting transition temperature, $T_C$, for two near-nodal cuts ($\theta_{FS}$=12$^\circ$ and 18$^\circ$) of a lightly underdoped ($T_C$=85K) Bi2212 sample. The raw spectra of Fig. 2a show the clear formation of the superconducting gaps as temperature is lowered, as well as the start of the ``bending back" Bogoliubov-type dispersion, which is most apparent in the coldest $\theta$=18$^\circ$ spectra. The corresponding TDoS are shown in panels 2b and 2c for the two cuts.  In both panels we see the TDoS taken above $T_C$ (i.e. the yellow 90K TDoS and the red 100K TDoS) still clearly show the presence of a finite gap by the depression of weight at $E_F$, even though the sample is no longer superconducting. The temperature dependence of $\Delta$ can be quantified by fitting the spectra with equation 1, the results of which are shown in Fig. 2d.  The angle dependence of $\Delta$ is consistent with a d-wave gap while the lack of any angle dependence of $\Gamma$ suggests the pair-breaking rate is isotropic.  The extracted $\Delta$ for each angle decreases only slightly from low to high temperature, including above the superconducting $T_C$. The superconducting gaps smoothly evolve through $T_C$ showing that the gaps we observe above $T_C$ have the same origin as the gaps below $T_C$ and are from pre-formed pairs.  Furthermore, we are well outside the region of the anti-nodal, and possibly competing, pseudogap which does not start until $\theta_{FS}=24^\circ$\cite{TanakaTwoGap}.

If we fit the observed temperature dependence of $\Delta$ to a conventional BCS form\cite{DahmLifetimes}, we find the gap closes at neither $T_C$ nor $T^*$, but at an intermediate temperature, $T_{Close}$ that is most consistent with recent experiments showing an onset of pairing in the intermediate region(Table \ref{tab:Temperatures})\cite{LuLiDiamagnetism,WangNernst,KondoSW}.  This $T_{Close}$ can be interpreted as a mean field temperature scale at which the pairing begins. However long range coherence and thus superconductivity does not occur until the lower temperature, $T_C$. We cannot yet resolve the gap as it completely closes so we cannot know the exact nature of the closing of the gap with temperature. It may even have a ``tail" extending to higher temperatures, as illustrated by the dashed orange lines in Fig. 2d.
\begin{table}
\begin{ruledtabular}
\begin{tabular}{ccccc}
Sample & $T_C$ & $T_{Close}$ & $T_{Nernst}$ & $T^*$ \\ \hline
 UD & 85K & $119\pm1$K & $130\pm10$K & 210K \\
 Opt & 91K & $114\pm2$K & $125\pm5$K & 170K \\
\end{tabular}
\end{ruledtabular}
\caption{\label{tab:Temperatures}Comparison of the temperature scales for the two dopings presented in this letter.  $T_C$ was determined from magnetometry, $T_{Close}$ as discussed, $T_{Nernst}$ from Wang et al.\cite{WangNernst}, and $T^*$ from the compilation by Hufner et al.\cite{Huefner2Gap} }
\end{table}

Our result is inconsistent with previous ARPES results that the near-nodal gap closes at $T_C$\cite{TanakaTwoGap}\cite{LeeTwoGap}. We argue that the difference is the failure of the old methods of gap determination to properly handle small gaps in the presence of large $\Gamma$ values. The most common method used today to determine the gap magnitude is the symmetrized EDC method, where $\Delta$ is typically determined by the peak location of the symmetrized EDC\cite{NormanFermiArc}.  In Fig. 2e, we show the symmetrized EDCs extracted from the same spectra that we extracted the TDoS in Fig. 2b.   At 90K (orange) the symmetrized EDC, though broadened, is not split into two peaks.  Under the traditional interpretation, this single peak means the gap has not opened.  However the TDoS at 90K clearly shows that $\Delta$ is still finite.  We find that this failure of the symmetrized EDC to resolve the finite $\Delta$ consistently occurs whenever $\Gamma > \Delta$(Fig. 2d). However, the TDoS can resolve the gap in this circumstance, as evidenced by the accuracy of fits for the TDoS extremely close to the node (Fig. 1d) and for the warm TDoS (Fig. 1e).

In the cuprates, the pseudogap is generally understood to be any gap that exists in the normal state, so our observation would qualify as a pseudogap.  However, most studies of the pseudogap have focused on the anti-node, where the gaps are largest, most easily measured and found to open at a high temperature $T^*$\cite{Hashimoto2}, not our intermediate $T_{Close}$.  Recent reports suggest that the anti-nodal pseudogap is likely due to ordering rather than pairing like the near-nodal pseudogap\cite{Hashimoto}.  Consequently, our results suggest the existence of two distinct pseudogaps: a pairing one observable near the node that exists up to $T_{Close}$ and an ordering-based one restricted to the anti-node (green on insets of Fig. 2b and 2c), that presumably exists up to $T*$.
\begin{figure}[htbp]
\includegraphics[width=90mm]{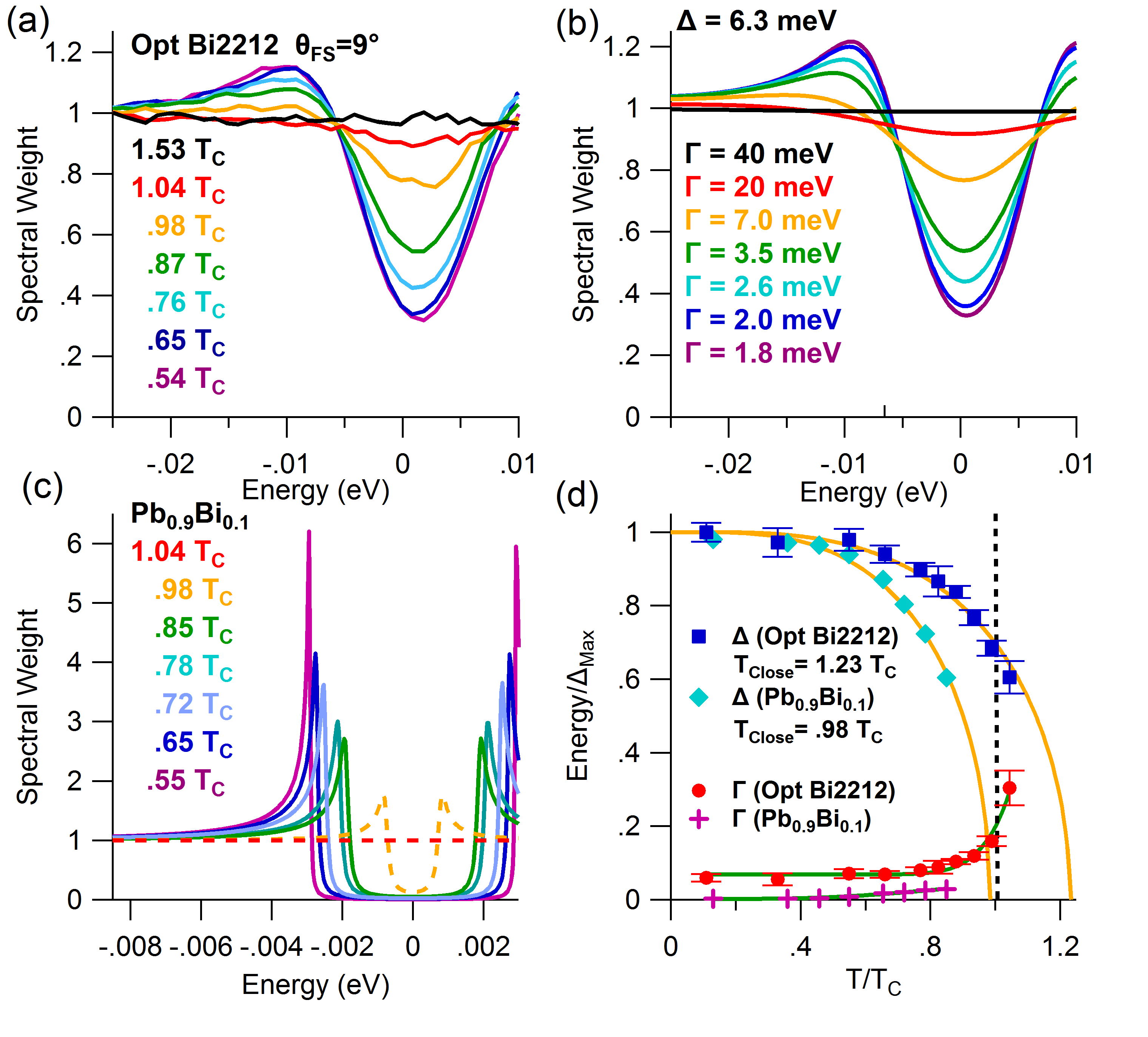}
\caption{\label{fig:Fig3}{\bf The `Filling' Gap of Bi2212} {\bf (a)} Temperature evolution of near nodal ($\theta_{FS}$ = 7.8$^\circ$) TDoS for optimally doped ($T_C$=91K) Bi2212.  {\bf (b)} Recreation of observed filling gap by only increasing the pair-breaking rate, $\Gamma$ while holding the pairing strength, $\Delta$ fixed {\bf (c)} Recreation of Dynes's original results showing the temperature dependence of the gap for a strongly coupled conventional superconductor.  Dashed curves are extrapolations based on the reported temperature dependence.  {\bf (d)} Comparison of fit results of Dynes formula to the original Pb$_{0.9}$Bi$_{0.1}$ to the TDoS of the optimally doped Bi2212 shown in panel A.  Here we have normalized both energy scales, $\Gamma$ and $\Delta$, to the maximum $\Delta$ as found at low temperature and, for Bi2212, the antinode. The error bars represent twice the statistical uncertainties($\pm 2\sigma$), as returned from our least-squares fitting procedures, to incorporate systematic errors as well.}
\label{fig3}
\end{figure}

As panels 2b and 2c show, the dominant effect of raising temperature is a shifting of weight from the peaks into the center of the gap, effectively ``filling" the gap, which has been observed by multiple spectroscopies  \cite{CampuzanoFilling,HwangOpticalScattering, BasovRMP,FischerRMP}. The fit results show that over our measured temperature range $\Delta$ has shrunk by 30\%, but $\Gamma$ has grown by 300\%.  We find the long-observed filling of the gap is due to this rapid and significant temperature evolution of $\Gamma$.  To illustrate this relation, Fig. 3a shows the temperature dependence of near nodal TDoS for an optimally doped sample of Bi2212.  Note that the gap is still finite in the normal state even for this optimally doped sample (red Fig. 3a).  Then, in Fig. 3b, we accurately recreate the observed temperature dependence using equation 1 by holding $\Delta$ fixed but varying $\Gamma$.  Perhaps the cleanest other work discussing gap filling and pair breaking is from STM studies across vortices in MgB$_2$\cite{EskildsenMgB2} and LuNiBC\cite{DeWilde}.  These studies show a similar filling of the superconducting gap attributed to enhanced pair-breaking from the increasing magnetic field.

For comparison to conventional BCS superconductivity, we consider the strongly-coupled superconductor Pb$_{0.9}$Bi$_{0.1}$ studied by Dynes\cite{DynesFormula}.  We recreated the original density of states from their reported values and equation 1.  For the warmest cases (dashed) we extrapolated the values of $\Gamma$ and $\Delta$. Even though the lead alloy material was chosen for its particularly strong pair-breaking rate, the dominant effect of raising temperature is still the closing of the gap, as evidenced by the shifting peaks.  As we present both the cuprate and BCS case with similar reduced temperature ($T/T_C$) scales it is easy to see how the `filling' of the gap in the cuprates is a fundamentally different evolution than the `closing' of the BCS gap.
To compare the interplay of $\Delta$ and $\Gamma$ for the BCS case and the cuprates, we present the temperature evolution of the two but normalized to the $\Delta_{Max}$: $\Delta(T=0)$ for the BCS case and $\Delta(T=0,\theta_{FS}=45^\circ)$ for the cuprates.  First, the cuprate $\Delta$ shows evidence for pre-pairing with a $T_{Close} \approx 1.23 T_C$, which is consistent with recent measurements of the Nernst effect\cite{WangNernst}, while the BCS case closes at $T_C$ as expected.  However, the pair-breaking rate is where the two materials diverge most dramatically.  For the BCS case $\Gamma/\Delta$ ranges from 0.5\% to no more than 5\%, even for temperatures very close to $T_C$, while the cuprates' $\Gamma$/$\Delta$ ranges from 6\% to 30\%. Thus the strength and temperature dependence of the pair-breaking processes in the cuprates are unusually strong, with significant consequences for the physics of high-$T_C$ superconductivity.

The results shown here indicate a qualitatively new type of superconductive pairing - not just the pre-formed pairs, but also the way the superconducting gap disappears as the sample temperature is raised. In a conventional BCS superconductor, $\Delta$ is reduced to zero as the temperature is raised to $T_C$.  In the cuprates, $\Delta$ does not close at $T_C$, but rather $\Gamma$ increases rapidly, filling in the gap such that it is mostly filled in when $T>T_C$.  The filling of the gap due to pair-breaking processes removes the energy gain for creating Cooper pairs, as opposed to the BCS mechanism where the gap magnitude itself shrinks. The temperature evolution of the superconducting energy gain for gapping is clearly different from the conventional BCS one and appears to have no clear precedent in other materials.  Further studies of the interplay of $\Delta$ and $\Gamma$ promise to reveal much about the cuprates and high-$T_C$ superconductivity. \nocite{KayNonlinearity,MannellaNonlinearity,ReberThesis,KordyukNL}

We thank A. Balatsky, I. Mazin, S. Golubov and M. Hermele for valuable conversations and D. H. Lu and R. G. Moore for help at SSRL. SSRL is operated by the DOE, Office of Basic Energy Sciences.  Funding for this research was provided by DOE Grant No. DE-FG02-03ER46066 (Colorado) and DE-AC02-98CH10886 (Brookhaven).
%
\end{document}